\documentclass[conference,letterpaper,9pt]{IEEEtran}
\pdfoutput=1

\usepackage[utf8]{inputenc}
\usepackage{pifont}

\usepackage{float}
\usepackage{varioref}

\usepackage[utf8]{inputenc}

\usepackage{alltt}
\usepackage{palatino}

\usepackage[english]{babel}
\usepackage[T1]{fontenc} 
\usepackage{verbatim} 
\usepackage{ifpdf}
\usepackage{tabularx} 
\usepackage{listings} 
\usepackage{graphicx}

\usepackage{algpseudocode}
\usepackage{rotating}
\usepackage{attachfile}
\usepackage{balance}
\usepackage{amsmath}
\usepackage{textpos}
\usepackage{numprint}
\usepackage{paralist}
\npthousandsep{,}
\usepackage{enumitem}
\usepackage{multirow}
\usepackage{framed}
\usepackage[usenames,dvipsnames]{xcolor}
\usepackage{lscape}
\usepackage{ltxtable}
\usepackage[]{mdframed}
\usepackage{tikz}
\usepackage{xcomment}

\usepackage{verbatim}
\usepackage{setspace}

\usepackage{amssymb}

\usepackage[figure,boxed,ruled,vlined]{algorithm2e}

\usepackage{subfigure}
\usepackage{xspace}
\usepackage{booktabs}

\usepackage{textcomp}

\usepackage[overload]{empheq}
\usepackage{letltxmacro}

\lstdefinestyle{mystyle}{
    keepspaces=true,                 
    numbers=left,                    
    showstringspaces=false,
    basicstyle=\footnotesize
}
\newlength{\figureskip}

\newlength{\betweentables}

\npthousandsep{,}
\npdecimalsign{.}

\newcommand*{\SavedLstInline}{}
\LetLtxMacro\SavedLstInline\lstinline
\DeclareRobustCommand*{\lstinline}{
  \ifmmode
    \let\SavedBGroup\bgroup
    \def\bgroup{
      \let\bgroup\SavedBGroup
      \hbox\bgroup
    }
  \fi
  \SavedLstInline
}

\definecolor{orange}{RGB}{255,127,0}
\definecolor{grey}{RGB}{135,135,135}

\newcommand\nbStrategies{9\xspace}
\newcommand\nbBugsInBenchmark{11\xspace}
\newcommand\nbBugsInBenchmarkRepairedAtLeastOnce{10\xspace}
\newcommand\nbseededfailures{519\xspace}
\newcommand\nbfixedseededfailures{318\xspace}

\newcommand{\tabincell}[2]{\begin{tabular}{@{}#1@{}}#2\end{tabular}}

\newcommand{\mycode}[1]{{\small \texttt{#1}}\xspace}

\newcommand\NPEFix{NPEfix\xspace}

\newcommand\stratReplaceVar{S1\xspace}
\newcommand\stratReplaceVarLocal{S1a\xspace}
\newcommand\stratReplaceVarGlobal{S1b\xspace}

\newcommand\stratReplaceNewLocal{S2a\xspace}
\newcommand\stratReplaceNewGlobal{S2b\xspace}
\newcommand\stratSkipLine{S3\xspace}
\newcommand\stratReturnNull{S4a\xspace}
\newcommand\stratReturnVar{S4c\xspace}
\newcommand\stratReturnNew{S4b\xspace}
\newcommand\stratReturnVoid{S4d\xspace}

\usepackage[]{pdfcomment}
\newcommand{\TODO}[1]{\textcolor{red}{#1}}\newcommand\todo\TODO

\begin{document}

\title{\NPEFix: Automatic Runtime Repair of Null Pointer Exceptions in Java}
\author{Benoit Cornu, Thomas Durieux, Lionel Seinturier and Martin Monperrus\\University of Lille \& Inria, France}
\maketitle

\label{chap:NPEFix}

\begin{abstract}
Null pointer exceptions, also known as null dereferences are the number one exceptions in the field.
In this paper, we propose \nbStrategies alternative execution semantics when a null pointer exception is about to happen.
We implement those alternative execution strategies using code transformation in a tool called \NPEFix.
We evaluate our prototype implementation on \nbBugsInBenchmark field null dereference bugs and \nbseededfailures seeded failures and show that \NPEFix is able to repair at runtime $\nbBugsInBenchmarkRepairedAtLeastOnce/\nbBugsInBenchmark$ and $\nbfixedseededfailures/\nbseededfailures$ failures.

\end{abstract}

\section{Introduction}
Null pointer exceptions, also known as null dereferences are the number one exceptions in the field \cite{li2006have}.
Li et al. found that 37.2\% of all memory errors in Mozilla and and Apache are null dereferences~\cite{li2006have}.
It is an inherent fragility of software in programming languages where null pointers are allowed, such as C and Java. 
A single null pointer can make a request or a task to fail, and in the worst case, can crash an entire application.

In essence, a null pointer exception is a violation of an assumption of the programmer: at this line, the variable about to be dereferenced is never null. 
However, in practice, the execution of a single line can result after a complex and long sequence of computation events, resulting in a violation of this assumption. Naturally, Kimura et al. \cite{Kimura2014returnnull} found that there are between one and four null checks per 100 lines of code on average.

There are two ways to combat null pointer exceptions. The first one is to forbid them upfront, and to ensure with static analysis than no null variable can ever be dereference. The second is to provide an alternative execution semantics when the null pointer exception happens. 
In this paper, we explore the second path. We define and evaluate \nbStrategies alternative execution semantics, which we call ``strategies''.

The strategies are categorized in two groups. The first group is about providing an alternative value when a null dereference is about to happen. 
This value can come from elsewhere in the memory (i.e. a valid value that is stored in another variable), or it can be manufactured on the fly. 
The second group of strategies is about skipping the execution of the null dereference. It can be either skipping a single statement or skipping the complete method.
All strategies are applicable for any arbitrary objects, incl.  instances of library classes, and instances of domain classes.

We implement those alternative execution strategies using code transformation. 
We design transformations that detect null dereferences just before the happen, and transformations that insert hooks for activating a given strategy at runtime. In our prototype implementation, the transformations are done on source code for Java programs. 
The prototype system is called \NPEFix. 

To evaluate \NPEFix, we perform two different experiments. 
First, we reproduce \nbBugsInBenchmark null pointer exceptions that happened in-the-field from open-source software and were reported in bug trackers. For each of those bugs, we transform the application code, run the crashing test case and see whether \NPEFix repairs the problem.
Second, we seed potential null dereferences at a larger scale. 
For three open-source projects, we remove all null-checks and run the test suite that comes with the project. The execution of the test suite involves a number of null values, which trigger null pointer exceptions due to the removed checks. 
In total, \NPEFix is able to automatically repair at runtime $\nbBugsInBenchmarkRepairedAtLeastOnce/\nbBugsInBenchmark$ field failures, and $\nbfixedseededfailures/\nbseededfailures$ seeded failures. 

To sum up, the contributions of this paper are:
\begin{itemize}
\item A set of \nbStrategies alternative execution strategies to repair null dereference failures.
\item A set of code transformations for allowing the configurable activation of those strategies at runtime.
\item \NPEFix, an implementation in Java of our technique that is publicly available on GitHub for supporting research on this topic.
\item The evaluation of \NPEFix on \nbBugsInBenchmark field null dereference bugs and \nbseededfailures seeded failures.
\end{itemize}

The remainder of this paper is organized as follows.
Section \ref{sec:concepts}  presents our approach for repairing null pointer exceptions at runtime.
Section \ref{sec:implementation} presents the implementation of of our approach in a tool called \NPEFix.
Section \ref{sec:eval} details the evaluation on \nbBugsInBenchmark field null dereference bugs and \nbseededfailures seeded failures.
Section \ref{sec:discussion} presents further analyzes and potential issues. 
Section \ref{sec:rw} presents the related works and Section \ref{sec:conclusion} concludes.

\section{Repairing Null Pointer Exceptions at Runtime}
\label{sec:concepts}
In this paper, we aim at automatically armoring software against crashes due to null dereferences.
Our technique, called \NPEFix, is based on two main steps: detecting potentially harmful null dereference at runtime, and providing an execution semantics alternative to crashing.

\subsection{Detecting Potentially Null Dereferences}
Our final goal is to repair null dereferences failures at runtime.
There are two main ways to repair a failure.
First, one can wait for the bug to happen, then deal with the error.
Second, one can act before the bug happens and avoid the error.
We decide to focus on the second way, i.e. to deal with the root cause of the bug.

Our detection of potentially harmful null dereferences is done in two steps.
First, we assess, each time a variable is going to be dereferenced, that this reference is null or not. This is done through additional checks added with source code transformation.

However, when a null dereference happens, it does not mean that there is a bug in the application.
In Java, when a null dereference occurs, an exception is thrown (e.g. \mycode{NullPointerException} in Java).
If this exception is caught at some point upper in the execution stack, it means that the application already contains a way to handle this error.
In other words, the dereference has been anticipated by the developers and cannot be considered as a bug.

On the opposite, when the exception thrown by the null dereference cannot be caught by the application, it means that the error is harmful, it will crash the application.
To detect those uncaughtable harmful null pointer exceptions, we build and maintain a runtime model of the try-catch of the stack, which will be discussed in Section~\ref{sec:baby-foot}.

\subsection{Repair Strategies}
\label{sec:strategies}
When an harmful null dereference is going to happen, there are two main ways to avoid it.
First, one can replace the null reference by a valid object, this way it will no longer be a \mycode{null} dereference.
Second, one can skip the problematic statement, and no null dereference will occur.
We refine those two techniques in \nbStrategies different strategies to repair null dereference failures.
A strategy is a set of actions which modify the behavior of the application in order to avoid the null dereference.

\newcommand*{\MyIndent}{\hspace*{0.8cm}}
\begin{table}
\begin{tabularx}{\columnwidth}{|l|l|l|l|X|}
\hline
\multicolumn{3}{|c|}{Strategy} &  Id & Description \\ \hline
\multirow{8}{*}{\rotatebox{90}{replacement~}} &\multirow{4}{*}{reuse} & local & \stratReplaceVarLocal & local injection of an existing compatible object \\ \cline{3-5}
&&global& \stratReplaceVarGlobal & global injection of an existing compatible object \\ \cline{2-5}
&\multirow{4}{*}{new} & local&\stratReplaceNewLocal & local injection of a new object \\ \cline{3-5}
&&global& \stratReplaceNewGlobal & global injection of a new object \\ \hline
\multirow{8}{*}{\rotatebox{90}{skipping}} & \multicolumn{2}{c|}{line}& \stratSkipLine & skip statement \\ \cline{2-5}
& \multirow{7}{*}{\rotatebox{90}{method}} & null & \stratReturnNull & return a null to caller \\ \cline{3-5}
&& new & \stratReturnNew & return a new object to caller \\ \cline{3-5}
&& reuse & \stratReturnVar & return an existing compatible object to caller \\ \cline{3-5}
&&  & \stratReturnVoid & return to caller (void method)\\ \cline{3-5}
\hline
\end{tabularx}
\caption{\NPEFix' alternative execution strategies upon null dereference failures.}
\label{tab:strategies}
\end{table}

\subsubsection{Strategies based on Null Replacement}
One way to avoid a null dereference to happen is to change the reference into a valid instance.
What object can be provided to replace the null reference?
We can inject an existing value (if one can be found) or a new value (if one can be constructed).
To facilitate the presentation we will use the following symbols.
\emph{r} is a reference and \emph{s} a statement dereferencing \emph{r}.
We basically want \emph{r} to reference a valid (non-null) value \emph{v}.

In both cases, the first thing to know is the required type.
In Java, this can be either a dynamic type of \emph{r} (if not-null) or a static type $T$ dereferenced by \emph{s}.
The fact that the program compiles implies that the type of \emph{r} is the same as, or a subtype of, the required type $T$.
So a valid replacement object is of compatible type, that is all subtypes of T.

Let us consider the case of injecting an existing object.
The set $S$ of the accessible objects is composed of the local variables, the method parameters, the class fields of the current class and all the other static variables.
This set is built and maintained at runtime thanks to code transformation.
Once $S$ is known, to obtain a valid object when a null dereference happens, we filter $S$ to only select  the set of all well typed and non-null values \emph{V}.
The two \stratReplaceVar strategies consists in testing all those values \emph{v in V} one by one.

Let us consider the case of creating a new value.
We statically know all the possible types for \emph{r}.
First, we filter those types to keep only the non-interface, non-abstract ones.
Then we try to create a new instance of each of those types (using reflection).
The two \stratReplaceVar strategies (\stratReplaceVarLocal and \stratReplaceVarGlobal) consist in testing all the successfully created instances one by one.

When a new object is provided  on the fly in place of the null dereference, it can be injected in the execution globally or locally.
\emph{Global injection} consists in providing an object (whether already existing or newly created) and assigning it to the null variable.
\emph{Local injection} consists in replacing one null reference to a valid object, without modifying the null variable itself.
This impacts the rest of the execution.
For global injection, all the other statements using \emph{r} will now perform their operations on \emph{v} instead of on \emph{null}.
For local injection, all the possible other statements using \emph{r} will still perform their operations on \emph{null}.
For local injection, it is likely that the new object \emph{v} is not stored anywhere else in the state after the execution of \emph{s} (except if it has been stored during \emph{s}, as a side-effect of its execution).

There are advantages and disadvantages for both kinds of injection.
The advantage of making a local modification (using \emph{r'}), is that we change as little as possible the state of the program.
But the corresponding disadvantage of a local modification is that we let a \emph{null} reference (\emph{r}) in the program, if \emph{r} is dereferenced again afterwards, we eventually have to do this choice again.
On the contrary, global injection permanently solves the nullity of $r$, but with a greater impact on the execution. 

This sums up in 4 possible strategies for the null replacement (see \autoref{tab:strategies}):
use an existing value locally (\stratReplaceVarLocal), use an existing value globally (\stratReplaceVarGlobal), use a new value locally (\stratReplaceNewLocal) and use a new value globally (\stratReplaceNewGlobal).

\subsubsection{Strategies based on Statement Skipping}
The second proposed way is to skip the statement where a null dereference would happen.
There are different possible skips.

The strategy \stratSkipLine consists in skipping the problematic statement and allows us to avoid the null dereference at this location.
We also propose a family of strategies which consists in skipping the rest of the method.
For skipping the rest of the method, there are two possibilities to consider.
Either the method returns nothing (\mycode{void}), and we can just return to the caller (strategy \stratReturnVoid), or the method expects a return value.
If the method expects a return value, we have to choose what to return to the caller and
we consider three possibilities.
First, we could return null, this is may look meaningless, but is actually a reasonable option because it is possible that the caller has a non-null check on the returned object.
Second, we could search in the set of the accessible values one which corresponds to the expected return type and return it.
Third, we could return a new instance of the expected type.
Those three strategies are respectively called \stratReturnNull, \stratReturnVar and \stratReturnNew.

All strategies are listed in \autoref{tab:strategies}.
The table represents the different dimensions of the analysis:
replacement vs skipping, local vs global, reusing objects vs creating new ones.
For each strategy, the corresponding code that needs to be injected is shown in the last column.

\subsection{Strategy Exploration}
\label{sec:strategy-exploration}

Now that we have defined alternative execution strategies, we need a way to explore them at runtime.
The algorithm of \autoref{fig:npe-algo} describes how strategy exploration is done within \NPEFix.
\NPEFix takes as input an application $A$.
First, it instruments $A$ with the transformations described in Section \ref{sec:implementation}.
Then, we assume that the application has a main loop, which is the case for most long running applications such as server-side applications, reactive applications and GUIs. 

When a dereference is about to happen, we first check that the exception will be caught using the runtime model of try-catch blocks.
If the exception will be caught, \NPEFix proceeds with the normal execution.
If the null pointer exception is about to crash the application, we perform the following actions.
First, \NPEFix selects the alternative strategies $S_p$ that have never been applied at this crash point $p$. 
They may be strategies which have never been used before at $p$, e.g. return instead of line skipping, or parametrized strategies with new parameters (e.g. return an object $x$ after the unsuccessful trial of returning $y$).
One strategy is then randomly picked in $S_p$.
Then, the strategy is applied and is marked as tried. This marking enables us to never try twice the same unsuccessful strategy.
If the strategy succeeds (no more errors happen), we deploy the strategy permanently at this crash point, which means that upon the next occurrences of the crashing null dereference, this valuable strategy will be automatically applied.
In addition, a patch is suggested to the developer.

Indeed, there is always a correspondence between a runtime alternative strategy and a source code patch.
For instance, let us consider strategy \stratReplaceVarGlobal, where the null variable $a$ is replaced at runtime by variable $b$.
This corresponds to the patch \mycode{if (a==null) {a.foo();} else {b.foo();}}.
A production version of \NPEFix would automatically send a patch email to a developer mailing list or automatically create a pull request on GitHub based on the successful application of a strategy.

\begin{figure}
\KwIn{Application $A$}
\Begin{
    instrument A\\
	\While{main loop\\}{
		\If{dereference is about to happen}{
			\If{the null pointer exception will be caught\\}{
				continue the execution
			}
			\Else{
				$S_p\gets$ untried strategies at crash point\\
				$s\gets $select randomly in $S_p$\\
				apply $s$\\
                          mark $s$ as tried\\
				\If{current task succeeds}{
					deploy the strategy permanently\\
                    suggest patch to developer
				}				
			}
		}
		
	}
}
\caption{The Main Algorithm of NpeFix.}
\label{fig:npe-algo}
\end{figure}

\section{Implementation}
\label{sec:implementation}

In this section, we present the implementation of the main different parts of the \NPEFix framework.

\subsection{Maintaining a Catch Stack at Runtime}
\label{sec:baby-foot}
At any code location during the execution, we want to know whether a null dereference may harm the application, 
where harmful is defined as the deference triggering an uncaught exception in the current thread (\mycode{NullPointerException} in Java).
In other words, we want to know whether the null pointer exception will be caught somewhere in the execution stack.
To do it, we use a code transformation that inject and maintain a stack of all exceptions that can be caught.
To know if a given exception will be caught at a given moment in the execution, we look at whether this exception type corresponds to one of the types in the stack.

The injection is done with program transformation. 
A source code transformation injects method calls at the beginning and the end of each try block and catch block. 
\autoref{fig:npe-stack} shows how this works in practice.
The method call to \mycode{catchStack.add} informs the framework that the execution enters in a try body which is able to catch the types given as parameter.
The method calls to \mycode{catchStack.remove} inform the framework that the executions exits the body of the given try.
At the end of every try, we remove the caught type from the stack.

There are three possibilities to exit the body of a try block:
1/ no exception is thrown: the end of the try is the end of the execution of the try (after codeA),
2/ a caught exception is thrown: the end of the try is the beginning of one catch (in the middle of codeA, just before codeB in \autoref{fig:npe-stack} ),
3/ an uncaught exception is thrown: the end of the try is the beginning of the finally block (in the middle of codeA).
To know when the try is finished we also add the call to \mycode{remove}  at the beginning of every catch and at the beginning of the finally.
In the cases 1/ and 3/, the \mycode{remove} call on the finally block allows us to know that the try is finished.
In the case 2/ the call on the corresponding catch block alerts the framework that the try is finished.
In this case, method \mycode{remove} is called twice, when the catch block is executed and in the finally block. This is not an issue, the framework knows that the try is no longer at the top of the stack and will not remove another try from the stack.

\begin{lstlisting}[numbers=none,caption={Code Transformation for Maintaining a Catch Stack at Runtime},label=fig:npe-stack,float]
// before transformation
try{
	// codeA
} catch (TypedException te){
	// codeB
}
// after NPEfix transformation
int tryId = catchStack.getFreeId();
try{
	catchStack.add(tryId, TypedException.class, AnotherTypedException.class);
	// codeA
} catch (TypedException te){
	catchStack.remove(tryId);
	// codeB
} finally {
	catchStack.remove(tryId);
}
\end{lstlisting}

\subsection{Detecting Null Dereference Before They Happen}

To detect null dereferences before they happen, we use a program transformation as follows.
We transform each method call and field access to a variable to insert a check. 
This is shown in \autoref{fig:npe-encaps}.

The call of \mycode{doSomething} that is originally present is now made on the result of method \mycode{checkForNull}.
Method \mycode{checkForNull} does the following things.
It first assesses whether the object is null, i.e. whether a null dereference will occur; if no, the program proceeds with its normal execution.
If the object is null, \NPEFix  looks at the try-catch stack presented in Section \ref{sec:baby-foot} to know whether the null pointer dereference is harmful.
If the upcoming null pointer exception will be caught, the program continues its normal execution.
If the exception will crash the execution, a runtime repair strategy is triggered. 

\begin{lstlisting}[numbers=none,caption={Detecting Harmful Null Dereferences With Code Transformation},label=fig:npe-encaps,float]
//before modification
o.doSomething();

// after NPEfix transformation
checkForNull(o).doSomething();

// with static method
public static Object checkForNull(Object o){
	if (o == null) // null dereference detected
		if (cannotCatchNPE())
			switch (STRATEGY) {
			  case s1b: return getVar(currentMethod());
			  ...
			}
	return o;
}
\end{lstlisting}

\subsection{Alternative Execution Strategies}
We have presented in Section \ref{sec:strategies} different strategies for repairing null dereference failures at runtime, we now present how we implement them.
They are all implemented using source code transformation.

\subsubsection{Value Replacement Strategies}
There are four strategies based on value replacement (the first half of \autoref{tab:strategies}):
\stratReplaceVarLocal, \stratReplaceVarGlobal, \stratReplaceNewLocal and \stratReplaceNewGlobal.

\begin{lstlisting}[numbers=none,caption={Maintaining a set of variables as pool for replacement at runtime}, label=fig:npe-getVar,float]
public void method(){
    ...
    Object a = {expr};
    a = {expr2};
    ...
}

public void method(){
    int id = getFreeId();
    startMethod(id, this);
    ...
    Object a = initVar({expr}, id, "a");
    a = modifVar({expr2}, id, "a");
    ...
    endMethod(id);
}
\end{lstlisting}

The first challenge is to maintain  set of variables as pool for replacement at runtime.
\autoref{fig:npe-getVar} shows how we tackle this problem; we use a stack to store all the variables of each method.
Each variable initialization and assignment inside the method is registered thanks to several calls to \mycode{initVar}.
In addition, at the beginning of the method, a call to \mycode{startMethod} uses reflection to access all the fields of the current instance.

In \autoref{fig:npe-encaps}, when an uncaughtable null pointer exception is about to happen, a call to method \mycode{getVar} looks for or creates a compatible object.
If the current activated strategy is a reuse-based replacement, the \NPEFix framework looks in the pool of objects a valid one and returns it.

Now, let's consider that the current strategy creates a new variable (strategies \stratReplaceNewLocal and \stratReplaceNewGlobal).
In this case, a call is made to \mycode{newVar}, this method takes as parameter the static type of the dereferenced variable.
\mycode{newVar} uses reflection to access to all the constructors of the given type.
In addition, this method is recursive so as to create complex objects if needed. 
Method \mycode{newVar} works as follows.
It tries to create a new instance of the class from each available constructor.
Given a constructor, it attempts to create a new instance for each of the parameter recursively.
The stopping condition is when a constructor does not need parameters.
Note that the primitive types, which don't have constructors, are also handled with a lower-level, straightforward machinery.

\subsubsection{Skipping Strategies}

Now we present how we implement the strategies based on skipping the execution (the second half of \autoref{tab:strategies}).

\paragraph{Statement skipping}
\begin{lstlisting}[numbers=none,caption={Implementation of Line-based Skipping},label=fig:npe-skipLine,float]
// before transformation
value.dereference(); 

// after NPEfix transformation
if (skipLine(value)){ 
    value.dereference();
}
boolean skipLine(Object... objs){ // NPEfix framework
    for (Object o : objs) {
        if (o == null && cannotCatchNPE() && doSkip())
            return false
    }
    return true;
}
\end{lstlisting}
The strategy \stratSkipLine necessitates to know if a null dereference will happen in a line, before the execution of the line.
For this, the transformation presented in \autoref{fig:npe-encaps} is not sufficient, because the call to method  \mycode{checkForNull} implies that the execution of the line has already started.
To overcome this issue, we employ an additional transformation presented in \autoref{fig:npe-skipLine}.

Similarly to \mycode{checkForNull}, method \mycode{skipLine} method assesses, before the line execution, if whether dereferenced value is null or not and whether it is harmful.
Method skipline takes an arbitrary number of objects, the ones that are dereferenced in the statement. This list is extracted statically.

There are numerous cases where one encapsulates a statement in an if-condition.
For example, one cannot skip a \mycode{return} or a \mycode{throw} statement, if the method has no corresponding return or throw in the other branch of the control-flow tree.
Also, it cannot skip variable declaration. 
To overcome this problem, we break the declaration and the first initialization, in order to only skip the first initialization.
All those cases are detected as non-skippable by our transformations.

\paragraph{Method skipping}
The remaining strategies are based on skipping the execution of the rest of the method, when a harmful dereference is about to happen: these are strategies \stratReturnVoid, \stratReturnNull, \stratReturnVar and \stratReturnNew (the last part of \autoref{tab:strategies}.
We implement those strategies with a code transformation as follows.

A try-catch block is added in all methods, wrapping the complete  method body.
This try-catch blocks handles a particular type of exception defined in our framework (ForceReturnError).
This exception is thrown by the \mycode{skipLine} method when one of the method-skipping strategies is activated, as show in \autoref{fig:npe-skipMeth}.
This listing also shows a minimalist example of the code resulting from this transformation.

\begin{lstlisting}[numbers=none,caption={Implementation of method-based skipping strategies},label=fig:npe-skipMeth,float]
// before transformation
Object method(){
	...
	value.dereference();
	...
	return X;
}
// after NPEfix transformation
Object method(){
	try {
		...
		if (skipLine(value)){
			value.dereference();
		}
		...
		return X;
	} catch (ForceReturnError f){
		if (s4a) return null;
		if (s4b) return getVar(Object.class);
		if (s4c) return newVar(Object.class);
	} 
}
boolean skipLine(Object... objs){
	if(hasNull(objs) && cannotCatchNPE() && skipMethodActivated())
		throw new ForceReturnError();
	...
}
\end{lstlisting}

\section{Evaluation}
\label{sec:eval}

We now present the evaluation of \NPEFix.
We evaluate in two ways.
First, we look at whether it is able to repair real world null dereferences.
Second, we create a large number of artificial null dereferences by seeding them into the code of open source research projects.

Our main research question is:\\
\newcommand\rqeffectiveness{RQ1: Is \NPEFix able to repair null dereference failures at runtime?}
\noindent\textbf{\rqeffectiveness}\medskip

To better characterize our system, we also answer to:\\ 

\newcommand\rqreuseorcreate{RQ2: What is the difference in effectiveness between strategies based on reusing objects vs based on creating new objects?}
\noindent\textbf{\rqreuseorcreate}\\

\newcommand\rqreplaceorskip{RQ3: What is the difference in effectiveness between null replacement and method skipping ?}
\noindent\textbf{\rqreplaceorskip}\\

\newcommand\rqoverhead{RQ4: What is the overhead of the system?}
\noindent\textbf{\rqoverhead}\\

\newcommand{\strategySucceeds}[1]{{\textbf{\color{LimeGreen}{$\bigstar$  #1}}}}
\newcommand{\strategyFailToBeApplied}[1]{{\color{red}{$\blacklozenge$  #1}}}
\newcommand{\strategyFailAfterapplication}[1]{{\color{blue}{$\blacksquare$ #1}}}
\newcommand{\mc}[2]{\multicolumn{#1}{c|}{#2}}

\begin{table*}
\centering
\begin{tabular}{|l|l|l|l|l|l|l|l|l|l|r|}\hline
 \multirow{4}{*}{Bug}  
 & \mc{4}{Replacement}             & \mc{5}{Skipping}            & \multirow{4}{*}{Success} \\\cline{2-10}
 & \mc{2}{Reuse}  & \mc{2}{New}    & \multirow{3}{*}{ \tabincell{c}{Line \\ \stratSkipLine}} & \mc{4}{Method}             &  \\\cline{2-5}\cline{7-10}
 & \tabincell{c}{Local  \\ \stratReplaceVarLocal}  & 
   \tabincell{c}{Global \\ \stratReplaceVarGlobal} & 
   \tabincell{c}{Local  \\ \stratReplaceNewLocal}  & 
   \tabincell{c}{Global \\ \stratReplaceNewGlobal} & 
          & 
   \tabincell{c}{null   \\ \stratReturnNull}       & 
   \tabincell{c}{New    \\ \stratReturnNew}        & 
   \tabincell{c}{Reuse  \\ \stratReturnVar}        & 
   \tabincell{c}{Void   \\ \stratReturnVoid}       & 
   \\
\hline
COLL-331 &\strategyFailToBeApplied{NoV}         & 
          \strategyFailToBeApplied{NoV}         & 
          \strategySucceeds{OK} &                 
          \strategySucceeds{OK} &                 
          \strategyFailToBeApplied{US}          & 
          \strategyFailAfterapplication{NPE}    & 
          \strategySucceeds{OK} &                 
          \strategySucceeds{OK} &                 
          \strategyFailToBeApplied{RI}          & 
          4 \\ \hline
LANG-304 & \strategyFailToBeApplied{NoV}        & 
          \strategyFailToBeApplied{NoV}         & 
          \strategySucceeds{OK} &                 
          \strategySucceeds{OK} &                 
          \strategyFailToBeApplied{US}          & 
          \strategyFailAfterapplication{NPE} &    
          \strategySucceeds{OK} &                 
          \strategySucceeds{OK} &                 
          \strategyFailToBeApplied{RI}          & 
          4 \\ \hline
LANG-587 & \strategyFailAfterapplication{NPE}   & 
          \strategyFailAfterapplication{NPE}    & 
          \strategyFailAfterapplication{NPE}    & 
          \strategyFailAfterapplication{NPE}    & 
          \strategySucceeds{OK} &                 
          \strategySucceeds{OK} &                 
          \strategySucceeds{OK} &                 
          \strategySucceeds{OK} &                 
          \strategyFailAfterapplication{NPE}    & 
          4 \\ \hline
LANG-703 & \strategyFailToBeApplied{NoV}        & 
          \strategyFailAfterapplication{NPE}    & 
          \strategySucceeds{OK} &                 
          \strategySucceeds{OK} &                 
          \strategySucceeds{OK} &                 
          \strategySucceeds{OK} &                 
          \strategySucceeds{OK} &                 
          \strategyFailToBeApplied{NoV}         & 
          \strategyFailToBeApplied{RI}          & 
          5 \\ \hline
MATH-290 & \strategyFailToBeApplied{NoV}        & 
          \strategyFailAfterapplication{NPE}    & 
          \strategyFailAfterapplication{Ex.}    & 
          \strategyFailAfterapplication{NPE}    & 
          \strategyFailToBeApplied{US}          & 
          \strategySucceeds{OK} &                 
          \strategySucceeds{OK} &                 
          \strategySucceeds{OK} &                 
          \strategySucceeds{OK} &                 
          4 \\ \hline
MATH-305 & \strategyFailToBeApplied{NoV}        & 
          \strategyFailToBeApplied{NoV}         & 
          \strategySucceeds{OK} &                 
          \strategySucceeds{OK} &                 
          \strategySucceeds{OK} &                 
          \strategyFailAfterapplication{Ex.}    & 
          \strategyFailAfterapplication{Ex.}    & 
          \strategyFailAfterapplication{Ex.}    & 
          \strategyFailAfterapplication{Ex.}    & 
          3 \\ \hline
MATH-369 & \strategyFailToBeApplied{NoV}        & 
          \strategyFailAfterapplication{NPE}    & 
          \strategySucceeds{OK} &                 
          \strategyFailAfterapplication{NPE}    & 
          \strategySucceeds{OK} &                 
          \strategyFailAfterapplication{NPE}    & 
          \strategySucceeds{OK} &                 
          \strategySucceeds{OK} &                 
          \strategyFailToBeApplied{RI}          & 
          4 \\ \hline
MATH-988a & \strategySucceeds{OK} &               
          \strategySucceeds{OK} &                 
          \strategySucceeds{OK} &                 
          \strategySucceeds{OK} &                 
          \strategyFailToBeApplied{US}            &   
          \strategyFailAfterapplication{NPE}  &   
          \strategySucceeds{OK} &                 
          \strategySucceeds{OK} &                 
          \strategyFailToBeApplied{RI}          & 
          6 \\ \hline
MATH-988b & \strategySucceeds{OK} &               
          \strategyFailAfterapplication{NPE}  &   
          \strategySucceeds{OK} &                 
          \strategyFailAfterapplication{NPE}  &   
          \strategyFailToBeApplied{US}        &   
          \strategyFailAfterapplication{NPE}  &   
          \strategySucceeds{OK} &                 
          \strategyFailToBeApplied{NoV}         & 
          \strategyFailToBeApplied{RI}          & 
          3 \\ \hline
MATH-1115 & \strategySucceeds{OK} &               
          \strategySucceeds{OK} &                 
          \strategyFailAfterapplication{Ex.}    & 
          \strategyFailAfterapplication{Ex.}    & 
          \strategySucceeds{OK} &                 
          \strategySucceeds{OK} &                 
          \strategyFailAfterapplication{Ex.}    & 
          \strategySucceeds{OK} &                 
          \strategyFailToBeApplied{RI}          & 
          5 \\ \hline
MATH-1117 & \strategyFailAfterapplication{Ex.}  & 
          \strategyFailAfterapplication{Ex.}    & 
          \strategyFailAfterapplication{Ex.}    & 
          \strategyFailAfterapplication{NPE}    & 
          \strategyFailAfterapplication{Ex.}    & 
          \strategyFailAfterapplication{Ex.}    & 
          \strategyFailToBeApplied{NoI}         & 
          \strategyFailAfterapplication{Ex.}    & 
          \strategyFailToBeApplied{RI}          & 
          0 \\ \hline\hline
 Total  & 3  
        & 2  
        & 7  
        & 5  
        & 5  
        & 4  
        & 8  
        & 7  
        & 1  
        & \nbBugsInBenchmarkRepairedAtLeastOnce  
        \\ \hline

\end{tabular}
\caption{The Effectiveness of NPEFix on 11 Field Null Pointer Failures.}
\label{tab:strategies-res}
\end{table*}

\subsection{Evaluation on Real Null References}
\label{sec:eval-real-bugs}

\textbf{RQ1a: Is \NPEFix able to repair field failures due to null dereferences?}

We collect and reproduce field failures due to null pointer exceptions, we apply our system to the buggy software, we see whether our system repairs the crashing dereference at runtime.

\subsubsection{Benchmark}
We build a benchmark of real null dereferences that happened in open-source projects.
It is based of the benchmark presented in the previous paper \cite{cornu:hal-01113988}.
It is composed of \nbBugsInBenchmark null dereference bugs.

There are two inclusion criteria.
First, the bug must be a real bug reported on a publicly-available forum (e.g. a bug tracker).
Second, the bug must be reproducible.
This point is crucial, it is really difficult to reproduce field failures, due to the absence of the exact crashing input, or the exact configuration (versions of dependencies, execution environment, etc.)

First, we look for bugs in the Apache Commons set of libraries (e.g. Apache Commons Lang).
The reasons are the following.
First, it is a well-known and well-used set of libraries.
Second, Apache commons bug trackers are public, easy to access and to be searched.
Finally, thanks to the strong software engineering discipline of the Apache foundation, a failing test case is often provided in the bug report.
To select the real bugs to be added to our benchmark we proceed as follows.
We took all the bugs from the Apache bug tracker\footnote{\url{https://issues.apache.org/jira/issues}}.
We then select 3 projects that are well used and well known (Collections, Lang and Math).
We add the condition that those bug reports must have \mycode{NullPointerException} (or \mycode{NPE}) in their title.
Then we filter them to keep only those which have been fixed and which are closed (our experimentation needs the patch).
Those filters let us 19 bug reports.
Sadly, on those 19 bug reports, 8 are not relevant for our experimentation:
3 are too old and no commit is attached (COLL-4, LANG-42 and Lang-144),
2 concern Javadoc (COLL-516 and MATH-466),
2 of them are not bugs at all (LANG-87 and MATH-467),
1 is about a VM problem.

Consequently, the benchmark contains \nbBugsInBenchmark cases from Apache Commons (1 from collections, 3 from lang and 7 from math).
This benchmark only contains real null dereference bugs and no artificial or toy bugs.
To reassure the reader about cherry-picking, we have not rejected a single reproducible field null  dereference.

\subsubsection{Results}

For each bug of the benchmark, we instrument the application code according to the source code transformations presented in Section \ref{sec:implementation}.
Then, we run each bug one by one by activating the strategies one after the other, it means that the experiment consists of \nbBugsInBenchmark $\times$ \nbStrategies = 99 executions.
\autoref{tab:strategies-res} presents the results of of this experiment.
Each line represents a null dereference bug,
each row represents the application of a strategy.

A cell with \strategySucceeds{OK} means that \NPEFix successfully repair the error at runtime, where successful means that no other exceptions are thrown after the application of the strategy.
A cell prefixed by a lozenge means that the strategy cannot be applied: \strategyFailToBeApplied{NoV} means that no valid object could be found for replacement-based strategies, \strategyFailToBeApplied{NoI} means that no object could created on the fly for object creation-based strategies, \strategyFailToBeApplied{RI} means that the strategy is incompatible with the return type of the method whether the null dereference happens (because the strategy returns \mycode{void} while the method expects an object or the other way around) and
\strategyFailToBeApplied{US} means that the null dereference happens on an unskippable line (e.g. a return with no control-flow counter part, as discussed above).
A cell prefixed by a rectangle means that a strategy could be applied but resulted in another error later in the execution flow, i.e. \NPEFix has replaced an error by another error.

The last column of the table is the number of viable strategies, if this number is higher or equals to 1, there is at least one effective strategy for this bug.
The last row shows the most effective strategy over all bugs.

There are 42 cases for which the application of a strategy repairs the bug (the number of table cells marked with \strategySucceeds{OK}).
In total, \nbBugsInBenchmarkRepairedAtLeastOnce out of \nbBugsInBenchmark null dereferences are repaired at runtime and 1 of \nbBugsInBenchmark are not because no strategy works at all.
On our benchmark.
We also see 10 bugs can  be repaired with several strategies (when the cell in the last column is higher than 1).
Recall that in the other case, our system simply results in the same exception as before, it means that our approach does not worsen the problem.

\begin{framed}
\NPEFix is able to repair at runtime \nbBugsInBenchmarkRepairedAtLeastOnce out of \nbBugsInBenchmark real null dereferences from our benchmark. 
\end{framed}

\subsection{Evaluation on Seeded Null Dereferences}
\label{sec:eval-seeded-bugs}

\textbf{RQ1b: Is \NPEFix able to repair failures due to seeded null dereferences?}

Now, we evaluate our approach by seeding null dereferences in existing open-source software. 
This is complementary to the evaluation on real bugs done in Section \ref{sec:eval-real-bugs}.
The previous evaluation is realistic but at a small scale.
On the contrary, by seeding null dereference, we trade realism for scale, because we are able to create many null dereferences.

\subsubsection{Protocol}

The evaluation protocol is as follows. 
First, we select open-source projects that come with test suites.
Second, we remove all null-checks in the application code (and not in the test code).
Third, we instrument the application source code with \NPEFix' transformations.
Fourth, we run the test suite of the application.

Since, all null checks have been removed from the application code, if they are null values involved in the execution of the test suite, this triggers several null dereferences.
This is actually the case for all applications we have looked at.
When a null deference is going to happen, it should be detected by \NPEFix, which then triggers one or several strategies.
This enables us to validate the effectiveness of \NPEFix' strategies on many dereferences.

\begin{table*}
\centering
\begin{tabularx}{\textwidth}{|X|r|r|r|r|r|r|}\hline
Project              & LOC     & \# tests & \# rem. checks & \# of failing NPE & \# of failing assert. & \# failures \\\hline
Spojo-core           & 993     & 135      & 13             & 87                & 2                  &   89 \\
Apache Commons Codec & 6876    & 694      & 4              & 51                & 10                 &   61 \\
Apache Commons Okio  & 3203    & 449      & 6              & 369               & 0                   &  369 \\ \hline
Total               & 11072       & 1278       & 23              & 497               & 12                  &  509\\
\hline

\end{tabularx}
\caption{Descriptive statistics of the \nbseededfailures  seeded failures}
\label{tab-desc-seeded}
\end{table*}

\autoref{tab-desc-seeded} gives the main statistics on those seeded bugs.
The first column is the project name, the second is its size in lines of code (LOC), the third is the number of removed null checks, the fourth is the number of failing test cases due to null pointer exceptions (NPE) without \NPEFix, the fourth is the number of test cases failing on assertions (also without \NPEFix).
For instance, project Spojo-core (the first row), we have removed 13 null checks, and this results in 87 test cases failing due to a null pointer exception (NPE) and 2 test cases failing due to a failing assertion.
In total, this protocol creates \nbseededfailures failures, which is an order of magnitude bugger than the reproduction of field failures. 

\subsubsection{Results}
\autoref{tab:npenullcheck} gives the results of this evaluation on seeded null dereferences.
Each row represents a project. 
The second column gives the total number of failures, and then, each column gives the number of repaired failures for each strategy (the higher the better).
For instance, for Spojo-core, we have seeded 89 failures, strategy \stratReplaceVarLocal repairs none of them while strategy \stratReturnVar automatically repairs $86/89$ of them.
The expected behaviour of the project is thus altered by \NPEFix.

Strategy \stratReplaceVarLocal (local injection of an existing object, third column) is unable to repair a single null dereference, while strategy \stratReturnVar (skip the method execution and returns an existing valid object) automatically repairs $86/89$ failures.
In total, the least effective strategies are \stratReplaceVarLocal and \stratReplaceVarGlobal (global and local injection of an existing object) and the most effective strategy is \stratReturnVar (skip the method execution and returns an existing valid object). Strategy \stratReturnVar alone is able to automatically repair at runtime
$283/\nbseededfailures$ (54,5\%) seeded failures.

\begin{framed}
\NPEFix is able to repair at runtime $\nbfixedseededfailures/\nbseededfailures$ (61\%) seeded null dereferences. 
\end{framed}

\subsection{Replacement Effectiveness}

\noindent\textbf{\rqreuseorcreate}\\

For providing a valid non-null object on the fly, we propose to techniques:
either we select a valid already existing object (strategies \stratReplaceVarLocal, \stratReplaceVarGlobal), or we create a new one (strategies \stratReplaceNewLocal and \stratReplaceNewGlobal).
We now discuss the respective effectiveness of those two techniques.

\paragraph{Replacement by an existing object}
For real null dereferences, as shown in \autoref{tab:strategies-res}, the strategies \stratReplaceVarLocal and \stratReplaceVarGlobal, which correspond to replacing the null value by a corresponding existing non null value, are not effective.
The reason is that in most of the studied cases no such value is available.
In all of those cases, the method is short and does not contain such a variable, and the class also does not contain a field of this type, or the field is null also.
For example, in  LANG-304, the required type is a Set, the method is 1 line-long, so there is no other values, and the only field of the class which is a Set is null.
This is confirmed for seeded bugs as well, as shown in \autoref{tab:npenullcheck}, where one sees that those two strategies are not effective at all: they do not repair a single seeded failure.

\paragraph{Replacement by a new object}
As shown in \autoref{tab:strategies-res}, strategies \stratReplaceNewLocal is able to repair $8/\nbBugsInBenchmark$ failures while \stratReplaceNewGlobal is able to repair five failures.
There are also cases where the replacement still results in an NPE later in the execution (e.g. for MATH-290 and \stratReplaceNewGlobal).
For the seeded failures, \autoref{tab:npenullcheck} gives a different picture.
global injection of new objects (\stratReplaceNewGlobal) is effective for project Okio and it is able to repair $159/369$ failures.

\begin{framed}
Injection of manufactured objects newly created on the fly is effective to repair null pointer exceptions.
\end{framed}

\subsection{Effectiveness of Skipping Strategies}
\noindent\textbf{\rqreplaceorskip}\\

Now we discuss in detail the effectiveness of the strategies based on skipping the execution of certain code regions.

\begin{table*}
\centering
\resizebox{\textwidth}{!}{
\begin{tabular}{|l|r||r|r|r|r|r|r|r|r|r|r|}\hline
\multirow{4}{*}{Project}  
 & \multirow{4}{*}{Failures}
 & \mc{4}{Replacement}             & \mc{5}{Skipping}            & \multirow{4}{*}{Union} \\\cline{3-11}
&& \mc{2}{Reuse}  & \mc{2}{New}    & \multirow{3}{*}{\tabincell{c}{Line \\ \stratSkipLine}} & \mc{4}{Method}             &  \\\cline{3-6}\cline{8-11}
& & \tabincell{c}{Local \\ \stratReplaceVarLocal}  & 
   \tabincell{c}{Global \\ \stratReplaceVarGlobal} & 
   \tabincell{c}{Local  \\ \stratReplaceNewLocal}  & 
   \tabincell{c}{Global \\ \stratReplaceNewGlobal} & 
          & 
   \tabincell{c}{null   \\ \stratReturnNull}       & 
   \tabincell{c}{New    \\ \stratReturnNew}        & 
   \tabincell{c}{Reuse  \\ \stratReturnVar}        & 
   \tabincell{c}{Void   \\ \stratReturnVoid}       & 
   \\
\hline
Spojo-core            & 89       & 0   & 0   & 1   & 0   & 88  & 86  & 82  & 82  & 82 & 89 \\
\small{Apache Commons Codec}& 61 & 0   & 0   & 0   & 0   & 22  & 6   & 0   & 0   & 0  & 23 \\
\small{Apache Commons Okio}& 369 & 0   & 0   & 11  & 159 & 0   & 12  & 0   & 201 & 0  & 206\\\hline
 Total                 & 519     & 0   & 0   & 12  & 159 & 110 & 104 & 82  & 283 & 82 & 318\\
\hline
\end{tabular}
}
\caption{Effectiveness of \NPEFix on the \nbseededfailures failures. Each cell gives the number of automatically repaired failures. The higher, the better.}
\label{tab:npenullcheck}
\end{table*}

\paragraph{Statement Skipping}
As shown in \autoref{tab:strategies-res}, the strategies \stratSkipLine, which corresponds to skipping the faulty line, is effective in 5 out of \nbBugsInBenchmark cases.
In 6 of the remaining failing cases, we can not skip the faulty line (US in the cell, meaning unskippable statement): 
four of them are variable initializations which are used later on (and putting an if not-null before the faulty line leads to a compilation error,as discussed in Section \ref{sec:disc-skip}.
The last one is a return statement, which also cannot be skipped for control-flow integrity.
There is only one case (MATH-1117) in our benchmark in which skipping the faulty line leads to another runtime error.
In this case, the skipped line is an assignment of a returned value, this value having already been assigned before.
Thus, the returned value later results in a badly constructed instance of \mycode{PolygonSet} (it has no \mycode{Plan}).
Finally, a method call on this bad instance of \mycode{PolygonSet} dereferences its Plan, which leads to another null dereference error.
In other terms, we replace one null dereference by another.
This is a good example of the difficulty to prevent every possible side effect when modifying the nominal behavior of an application.
For the seeded failures of \autoref{tab:npenullcheck}, \stratSkipLine is quite effective for projects Spojo-Core and Commons-Codec with resp. $88/89$ and $22/61$ repaired failures.
This is normal since  skipping the line is the closest equivalent repair to the seeded strategy (removing null-checks).

\paragraph{Method Skipping}
Now we consider the strategies that consists of returning from the current method if a null dereference is about to happen (strategies \stratReturnVoid, \stratReturnNull, \stratReturnVar and \stratReturnNew).
There are two cases in our benchmark where the null dereference occurs inside a method which returns void (MATH-290 and MATH-305).
Those cases correspond to the strategy \stratReturnVoid.
In MATH-290, we are able to return instead of executing the faulty statement, and the test case passes.
Skipping the method cancels the execution of the \mycode{initialize} method called at the end of the \mycode{SimplexTableau} constructor. 
However, a valid not-null object is returned, as expected by the test case.
For MATH-305, skipping the \mycode{assignPointsToClusters} method leads to a division by zero later in the program execution.
This method was clearly essential, and skipping its whole execution is not a valid substitute.
Interestingly, in this case, skipping the faulty line (\stratSkipLine) is a valid substitute, which means that this particular line in the method is not essential, the essential part being in the rest of the method body.
For the seeded failures of \autoref{tab:npenullcheck}, \stratReturnVoid is quite effective for projects Spojo-Core with $82/89$ repaired failures.
This is normal since most of the failures appear in void methods and the test suite does not have strong assert on the output .

There are 9 cases in our benchmark where the null dereference occurs inside a method which returns something.
Those cases can be handled by strategies \stratReturnNull, \stratReturnVar and \stratReturnNew.
Let us first consider the degenerated case (strategy, \stratReturnNull, that returns null in the presence of a null dereference).
As expected, this strategy leads to another errors (null dereference in 5 cases and two runtime errors in the remaining cases).
Those errors are all consequences of the injected null value.
Surprisingly, this strategy is a valid strategy in 4 out of \nbBugsInBenchmark cases.
For instance, for MATH-290 the method takes an array as input and creates another array which corresponds to the types of the values in the parameter array.
Returning a null array instead of a array containing a null value.
This validates the hypothesis presented in Section \ref{sec:concepts} that often, the caller has a non-null check on the returned object.
For seeded bugs, surprisingly, returning is extremely effective for the project Spojo-core, because almost all failures appears in void method (82) and only 4 failures are repaired with a returns null.
This strategy also fixes 13 failures in the project Apache Commons Codec (1 failure) and Apache Commons Okio (12 failures) because they have test cases that attends null pointer exceptions.
Although since all null-check have been removed, our experimental protocol with seeded bugs is irrelevant for this particular strategy.

\autoref{tab:strategies-res} also shows that returning an existing value of the expected return type (\stratReturnVar) is a valid strategy in 7 out of the \nbBugsInBenchmark cases.
We encounter the same problems as replacing the dereferenced value by an existing one, in 2 out of the \nbBugsInBenchmark cases, no corresponding value can be found.

Returning a new value of the expected return type (\stratReturnNew) is a valid strategy in 8 out of the \nbBugsInBenchmark cases, which is the best performance according to this benchmark.
This result is wholeheartedly confirmed by the experimental results on seeded failures. Strategy \stratReturnNew is able to automatically repair $82/\nbseededfailures$ seeded failures at runtime.

\begin{framed}
When a null dereference is about to happen, skipping the rest of the method body being executed is effective, both on field and seeded failures. This alternative execution strategy is able to successfully repair $9/\nbBugsInBenchmark$ field failures and $201/\nbseededfailures$  seeded failures.
\end{framed}

\subsection{Overhead}
\label{sec:overhead}
\noindent\textbf{\rqoverhead}\\

\begin{table}
\centering
\begin{tabular}{|l|r|r|r|}\hline
\multirow{2}{*}{Project} & \multicolumn{2}{c|}{Execution time (ms)}    & \multirow{2}{*}{Overhead} \\\cline{2-3}
                         & original   & after trans.         &       \\
\hline
 Spojo-core          & 336            & 381                  & 13\%     \\
 Apache Codec        & 25885          & 29090                & 12\%     \\
 Apache Okio         & 9857           & 15657                & 58\%     \\
\hline

\end{tabular}
\caption{Comparison of execution on the original and transformed code by \NPEFix. Absolute time are given in milliseconds.}
\label{tab:overhead}
\end{table}

Now, we analyze the runtime overhead introduced by the code transformations applied by \NPEFix.
For each project of our dataset, we execute 10 times the test suite on the original application code, and then 10 times the same test suite on the transformed application code. 
\autoref{tab:overhead} gives the result of this experiment.
For each project, it gives the execution time of a single execution of the test suite before and after transformation, and the corresponding overhead in percentage.
For instance, for Spojo-core, the test suite runs in 336 milliseconds originally and in 381ms after transformation. 
In total, the overhead is quite variable, it ranges from 12 to 58\%. The maximum value is for Apache Commons Okio, because the execution time increases considerably (ten times slower) in two test cases of this project.
These two test cases call an important amount of loop iterations.

\section{Discussion}
\label{sec:discussion}
\subsection{Threats to Validity}
A bug in our system is the biggest threat to the validity of our results. We have made our system publicly available on GitHub for other researchers to reproduce our findings.
There are several threats to the external validity:
first the field failures we were able to reproduce might not represent the diversity and distribution of null dereferences that happen in the field;
second, as for all experiments with seeded bugs, there is no warranty that the seeded bugs are realistic (although we are quite confident in this case, since many real-world patches consist of adding non-null checks);
third, our findings on the Java programming language might not hold for other languages suffering from null dereferences.

\subsection{Limitations}
\subsubsection{Null Dereference Location}
Because our approach uses source code modification, we have to have access to the source code of the location of the null dereference.
If the dereference happens in an archived library, our approach is not able to detect the null dereference.

\subsubsection{Unskippable Line}
\label{sec:disc-skip}
In several cases, it is impossible to add a guard before a statement.
For example, if the statement is a variable declaration and/or initialization, one  cannot skip it because other locations may use the declared variable.
Also it is problematic if the null dereference happens in an expression inside a condition.
For example, what does it mean to skip the execution of an if condition? 
A similar problem happens for null dereferences in loop conditions.

\section{Related Work}
\label{sec:rw}

There are several static techniques to find possible null dereference bugs.
Hovemeyer et al.~\cite{hovemeyer2005evaluating} use byte-code analysis to provide possible locations where null dereference may happen.
Sinha et al.~\cite{sinha2009fault} use source code path finding to find the locations where a bug may happen and apply the technique to localize Java null pointer exceptions symptom location.
Spoto \cite{spoto2011precise} devises an abstract interpretation dedicated to null dereferences \cite{spoto2011precise}.
Ayewah and Pugh \cite{ayewah2010null} discussed the problems of null dereference warnings that are false positives.
Compared to these works, our approach is  dynamic and instead of predicting potential bugs, it repairs the actual ones that result to crashes in production .

Romano et al. \cite{romano2011approach} find possible locations of null dereferences by running a genetic algorithm to exercise the software.
If one is found, a test case that demonstrate the null dereference is provided.
The Linux kernel employs special values, called poison pointers, to transform certain latent null errors into fail-fast errors \cite{rubini2001linux}.
Contrary to our approach, poison values only provide fail-fast behavior and do not repair the problem at runtime.
Bond et al.~\cite{bond2007tracking} present an approach to dynamically provide information about the root cause of a null dereference (i.e. the line of the first null assignment). 
None of those dynamic techniques perform repair as we do.

Yong et al. \cite{yong2003protecting} use static analysis to detect potential null dereferences and inject runtime checks.
The runtime checks capture unsafe dereference and log them as potential security violations, before halting the execution. 
This is not runtime repair as we mean in this paper.

Jeffrey et al. \cite{jeffrey2010execution} have proposed a diagnostic algorithm to find the root cause of memory errors.
Their idea is to study the next crash after suppressing the original one.
The key difference is their approach is a diagnosis one and not a repair one.
However, the common point is that they study the next error when the original one is masked.
This is also what we do indirectly, when we observe in \autoref{tab:strategies-res} that the original null pointer exceptions are replaced by other runtime errors later one, which means that the system state is remains unstable.

Lin et al. \cite{lin2007autopag} tries to generate a source code patch from a working exploit that triggers an array overflow in C code.
Its repair operators consist of fixing out-of-bound reads by adding a modulo in the read expression and out-of-bound writes by truncating data to be written.
The error model (array overflow) is different from ours (null deferences).

Assure \cite{sidiroglou2009assure} is a self-healing system based on checkpointing and error virtualization.
Error virtualization consists of handling an unknown and unrecoverable error with error-handling code that is already present in the system yet designed for handling other errors.
In Assure, error virtualization is associated with fuzzing to discover and test in advance valuable error virtualization points, called rescue points, which correspond to the hooks we insert for repairing null pointer exceptions.

Dobolyi and Weimer~\cite{dobolyi2008changing} present a technique to tolerate null dereferences.
Using code transformation, they introduce hooks to a recovery framework.
This framework is responsible for forward recovery of the form of creating a default object of an appropriate type to replace the null value or of skipping instructions.
The key differences with our work is that they do not explore replacement, local and global injection, method level skipping, and the evaluation is much smaller.

Kent~\cite{kent2008dynamic} also proposes alternatives to null pointer exceptions.
The skip and return and new object strategies are directly inspired from Kent's work.
However, the idea of variable replacement is new, as well as the exploration of local versus global injection.
Our empirical results complement theirs by providing new insights about the ability of the existing and new strategies to repair failures from a different benchmark.

Recently, Long et al. \cite{LongSR14} have introduced the idea of ``recovery shepherding''.
Upon null dereferences, recovery shepherding consists in returning a manufactured value and to track it during the execution.
While their work target simple manufactured value of primitive datatypes in C, our work is in Java, where we reason about complex abstract data types (for instance by creating instances on the fly). Also, the replacement-based techniques and local injection are novel techniques not explored in \cite{LongSR14}.

\section{Conclusion}
\label{sec:conclusion}

In this paper, we have presented \NPEFix, a novel approach for repairing harmful null dereferences at runtime.
We proposed a set of \nbStrategies alternative execution strategies that are able to repair this type of failure at runtime.
We have presented code transformations 1) to detect the harmful null dereferences at runtime; 2) to allow a behavioral modification for executing the strategies.
\NPEFix is able to repair at runtime \nbBugsInBenchmarkRepairedAtLeastOnce out of \nbBugsInBenchmark real null dereferences from our benchmark. 
In an evaluation with seeded bugs, we have found that \NPEFix is able to repair \nbfixedseededfailures out of \nbseededfailures seeded null dereference failures. 
When a null dereference is about to happen, skipping the rest of the method body being executed is the most effective according to our experiments.
This alternative execution strategy is able to successfully repair $9/\nbBugsInBenchmark$ field failures and $210/\nbseededfailures$ seeded failures.

\bibliographystyle{abbrv}
\bibliography{references}

\end{document}